\documentclass[10pt,letterpaper]{article}

\newcommand{\m}{\mathbf}

\newcommand{\lt}{\left<}
\newcommand{\rt}{\right>}

\usepackage{opex3}
\usepackage{graphicx}
\usepackage{amsmath}
\usepackage{verbatim} 
\usepackage{color}

\begin{document}

\title{ Design of Optomechanical Cavities and Waveguides on a Simultaneous Bandgap Phononic-Photonic Crystal Slab}

\author{Amir H. Safavi-Naeini and Oskar  Painter}

\address{Thomas J. Watson, Sr, Laboratory of Applied Physics, California Institute of Technology, 
Pasadena, California 91125, USA }

\email{safavi@caltech.edu}
\email{opainter@caltech.edu} 
\homepage{http://copilot.caltech.edu}


\begin{abstract}
  In this paper we study and design quasi-2D optomechanical crystals, waveguides, and resonant cavities formed from
  patterned slabs. Two-dimensional periodicity allows for in-plane pseudo-bandgaps in frequency where resonant optical
  and mechanical excitations localized to the slab are forbidden. By tailoring the unit cell geometry, we show that it
  is possible to have a slab crystal with simultaneous optical and mechanical pseudo-bandgaps, and for which optical
  waveguiding is not compromised. We then use these crystals to design optomechanical cavities in which strongly
  interacting, co-localized photonic-phononic resonances occur.  A resonant cavity structure formed by perturbing a
  ``linear defect'' waveguide of optical and acoustic waves in a silicon optomechanical crystal slab is shown to support
  an optical resonance at wavelength $\lambda_0 \approx 1.5 ~\mu\text{m}$ and a mechanical resonance of frequency
  $\omega_m/2\pi \approx 9.5 ~\text{GHz}$.  These resonances, due to the simultaneous pseudo-bandgap of the waveguide
  structure, are simulated to have optical and mechanical radiation-limited $Q$-factors greater than $10^7$. The optomechanical coupling of the optical and acoustic
  resonances in this cavity due to radiation pressure is also studied, with a quantum conversion rate, corresponding to the scattering rate of a single cavity photon via a single cavity phonon, calculated to be $g/{2\pi} = 292 ~\text{kHz}$.
\end{abstract}

\ocis{(000.0000) General.} 





\section{Introduction}

Recent efforts~\cite{Kippenberg2008,Favero2009} to utilize radiation pressure forces to probe and manipulate micro- and nano-scale mechanical objects has spawned a number of new types of optomechanical systems.  Amongst these, guided wave nanostructures in which large gradients in the optical intensity are manifest, have been shown to possess extremely large radiation pressure effects~\cite{Chan2009,Eichenfield2009a,Li2008,Lin2009,Wiederhecker2009,Roels2009}.  The recent demonstration of strongly interacting co-localized photonic and phononic resonances in a quasi one-dimensional (1D) optomechanical crystal (OMC)~\cite{Eichenfield2009c, Eichenfield2009d} has shown that it may be possible to coherently control phonons, photons, and their interactions on an integrated, chip-scale platform. Considering the already wide applicability of quasi two-dimensional (2D) slab photonic crystals~\cite{Painter1999,Painter1999a,Vuckovic2001,Takano2006,Song2005a,Notomi2005} and 2D phononic crystals~\cite{OlssonIII2009,Sanchez-Perez1998,Vasseur2007,Gu2006,Mohammadi2008a}, and for purposes of integration and obtaining better optical and mechanical mode localization, it is desirable to investigate the prospects of a quasi-2D OMC architecture.

Previous studies of the photonic and phononic properties of infinitely thick 2D crystal structures have shown, amongst other things, the practicality of deaf and blind structures with simultaneous photonic and phononic in-plane bandgaps~\cite{Maldovan2006a}, the experimental demonstration of a full in-plane phononic bandgap~\cite{Gorishnyy2005}, and the theoretical prospects of strongly co-localizing light and sound in defect cavities~\cite{Maldovan2006}.  More recently, studies of quasi-2D silicon slab structures have been performed in which phononic bandgap mirrors have been experimentally measured~\cite{Mohammadi2009} and crystal structures possessing a simultaneous in-plane bandgap for guided photons and phonons have been proposed~\cite{Mohammadi2008,Mohammadi2010}.  Each of these past studies was performed in the linear regime, neglecting photon-phonon interactions resulting from radiation pressure.  Radiation pressure effects have previously been investigated theoretically in quasi-2D photonic crystal structures~\cite{Notomi2006}, albeit in double-layer slab structures in which it is the flexural mechanical modes of the slabs which give rise to the optomechanical interaction.  Perhaps closer to the subject of our work is the recent study of optomechanical interactions in quasi-2D photonic crystal fibers~\cite{Laude2005,Dainese2006a,Brenn2009,Kang2009}, in which measurements have shown that these crystal fiber systems can support traveling photonic and phononic modes with a strong nonlinear optical interaction. 

In this article we aim to build on our previous work with quasi-1D nanobeam OMCs, and develop quasi-2D thin-film crystal structures capable of efficient routing, localization, and interaction of light and sound waves over the full plane.  Unlike in the infinitely thick 2D crystal structures, thin-film structures require certain limits on the air filling fraction in order to maintain effective optical waveguiding, thus making it more challenging to realize a phononic bandgap.  We begin in Section \ref{sec:crystal} with the study of two new crystal structures for this purpose, one called the ``cross'' substrate and the other the ``snowflake'' substrate.  These thin-film crystal structures possess large in-plane phononic bandgaps by incorporating resonant elements in the substrate~\cite{Leroy2009}. After investigating the origin of these large bandgaps in varying levels of detail, we move to studying the photonic properties of these structures, and demonstrate in Section \ref{ss:snowflake} that the ``snowflake'' substrate possesses a large simultaneous optical bandgap. From here we move in Section \ref{sec:waveguide} to creating one-dimensional defects, i.e. waveguides, in the ``snowflake'' structure, and studying their tuning properties. In Section \ref{ss:WG_coupling}, to understand the optomechanical coupling, we also study the guided-mode optomechanical coupling in these  waveguide structures and find their simple relation to the expected cavity optomechanical coupling. We culminate our analysis in Section \ref{sec:cavity} with the design of an optomechanical cavity on a silicon slab with an optical $Q > 5\times 10^7$ and an extremely large optomechanical coupling rate.

\section{Crystal Design}\label{sec:crystal}

In this section we study two types of crystal structures, one with a square lattice and the other with a hexagonal lattice. Most photonic and phononic crystals to date have utilized circular holes. Here we investigate the degree of freedom associated with hole shape. Circular holes are the simplest holes to fabricate and the most symmetric. Considering that it is desirable for the hole to have at least as much (point-group) symmetry as the underlying lattice, circular holes seem like the obvious choice. Unfortunately, circular holes fail to provide large mechanical bandgaps at desirable frequencies, and fail completely at providing  a bandgap for a variety of slab thicknesses. In cases where full simultaneous phononic-photonic bandgaps are achievable, e.g. the square lattice proposed by Mohammadi, et al., in Refs.~\cite{Mohammadi2008,Mohammadi2010}, the large hole sizes make the crystal unsuitable for fabrication of ultrahigh-$Q$ optical cavities. Thankfully, by exploring the shape degree of freedom of the hole it is possible to do better when it comes to phononic bandgap materials.

\subsection{Origin of the Gap: From Effective Medium to Tight-Binding}\label{ss:tight_bind}

\definecolor{MyDarkBlue}{rgb}{0.1,0,0.55}
\definecolor{MyCyan}{rgb}{0.1,0.6,0.7}
\definecolor{MyGreen}{rgb}{0.1,0.6,0.1}
\definecolor{MyRed}{rgb}{0.8,0.1,0.1}

\begin{figure}[ht]
\begin{center}
\includegraphics[width=0.85\columnwidth]{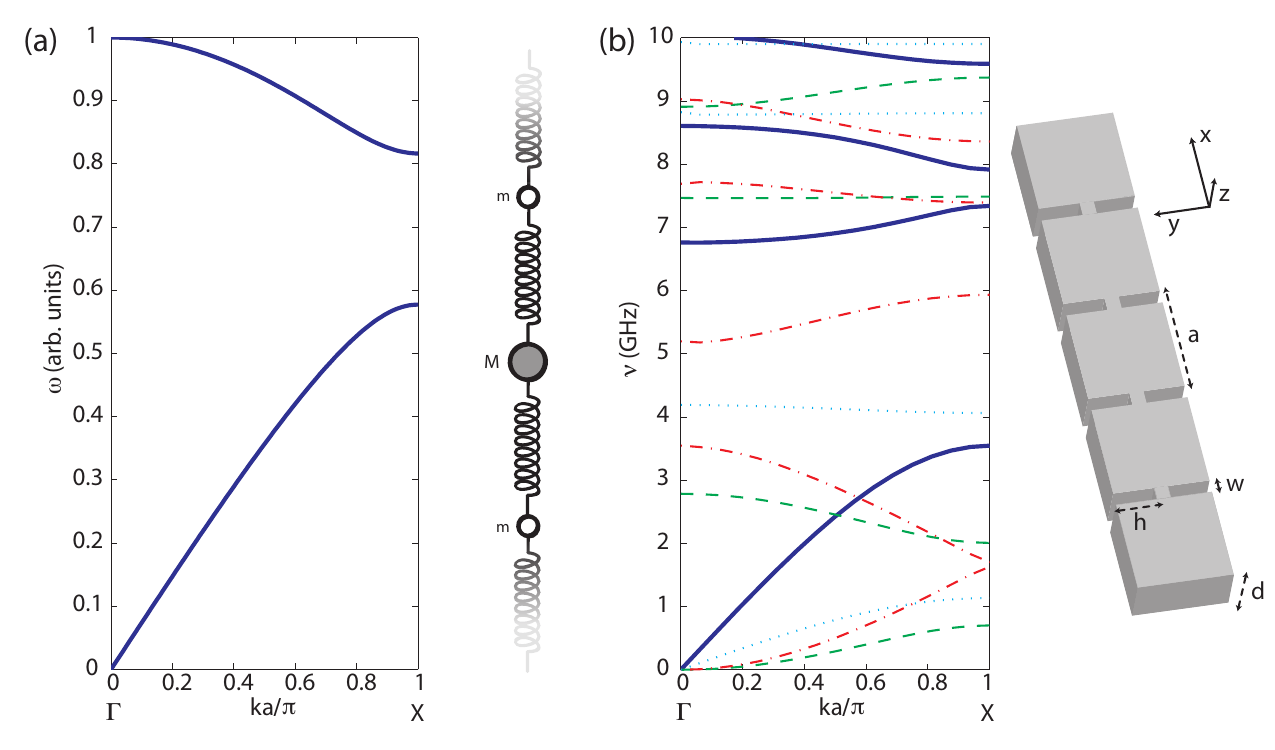}
\caption{Band diagram for a simple (a) spring-mass system and (b) its quasi-1D nanomechanical analogue. The nanomechanical structure band diagram is plotted for a silicon structure with $(d,h,w,a)=(220,200,100,500)~\text{nm}$. In (b) waveguide bands are plotted as ({\color{MyDarkBlue}---}), ({\color{MyCyan}$\cdot \cdot \cdot$}), ({\color{MyGreen}-~-~-}) and ({\color{MyRed}$\cdot - \cdot$}) for vibrational modes with $(\sigma_y,\sigma_z)$ symmetry of $(+,+)$, $(-,-)$, $(-,+)$ and $(+,-)$ respectively.  Here, $\sigma_y$ is a mirror plane symmetry of the structure about the $y$-axis.  $\sigma_z$ is the corresponding vertical mirror symmetry about the $z$-axis.}
\label{fig:1dcrystal}
\end{center}
\end{figure}

 We start with the toy model consisting of alternating small and large masses coupled by springs typically studied in solid-state physics~\cite{Kittel2005} . As shown in Figure~\ref{fig:1dcrystal}(a), this system results in a large bandgap in the frequency spectrum of the mechanical vibrations of this linear chain. The bandgap arises from the existence of modes of oscillation with widely different frequencies. Traveling phonon modes where the large masses are predominantly excited (low frequency ``acoustic''-like phonons) are split in frequency from those where the small masses are predominantly excited (high frequency ``optical''-like phonons). The frequency bandgap between low and high frequency vibrations becomes more pronounced with increasing difference in particle masses. A quasi-1D nanomechanical realization of the linear chain of alternating small and large masses coupled by springs is shown in Figure~\ref{fig:1dcrystal}(b).  This structure consists of a linear array of square ``drumheads'' joined together by narrower ``connector'' pieces. Here, and throughout the rest of this article, the material of the mechanical structure is assumed to be silicon (Si), with Young's modulus of $170$ GPa and mass density of $2329$ kg/m$^3$.  These and the following mechanical simulations were done using COMSOL Multiphysics \cite{COMSOL2009}, a finite-element solver. We see that large phononic bandgaps exist in this structure just as in the elementary periodic array of small and large masses coupled via springs. 

To more accurately understand the origin of the bandgap in the quasi-1D nanomechanical structure of Figure~\ref{fig:1dcrystal}(b), we consider the low and high frequency bands separately.  In the low-frequency, small wavelength ($\lambda \ll a$) regime, an effective medium theory \cite{Nemat-Nasser1993,Mei1996} can be used to calculate the exact values of wave velocities and the dispersion at the origin. For our purposes, we wish to push these low frequency bands to as low a frequency as possible. This can be achieved by reducing the width of the connector piece, $a-2h$. The reasoning heuristically is as follows.  The stiffness of a beam is in a sense dependent on the stiffness of its weakest link. In this case, the connectors, acting as contacts between the larger square drumhead sections, make the beam much floppier than an unperturbed beam of uniform cross-section. At higher frequencies, as $\nu$ approaches the resonances, $\nu_j$, of each square drumhead, a tight-binding model may be used to model the dispersion. This sort of model is valid as long as the interaction between each square is made small, which we have achieved by making $a-2h$ small.  To get a bandgap then, the effective medium bands are squeezed down to low frequencies by reducing $a-2h$, which does not affect the tight-binding bands since their frequency is set by the internal resonances of the larger square drumhead section. Simultaneously, the interaction strength between the coupled drumhead sections is reduced, and therefore so is the slope of the tight-binding bands making them more flat. These two effects conspire together to produce a large phononic bandgap.

\subsection{Quasi-1D Phononic Tight-Binding Bands: Symmetry and Dispersion}\label{ss:slopes}

\begin{figure}[htbp]
\begin{center}
\includegraphics[width=0.85\columnwidth]{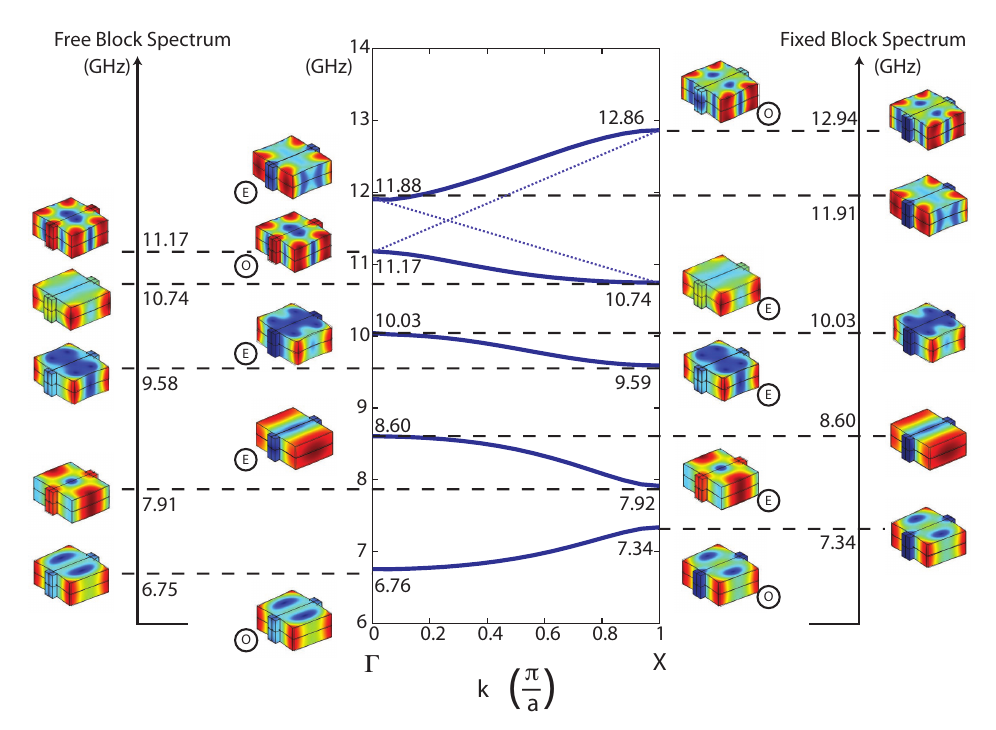}
\caption{The tight-binding bands of symmetry $(\sigma_y,\sigma_z) = (+,+)$ are plotted for the quasi-1D nanomechanical system of Fig.~\ref{fig:1dcrystal}(b), along with the unit cell solved for fixed and free boundary conditions. The $E$ ($O$) symbol below the Bloch function plots designates $\sigma_x = + (-)$ symmetry of the vibrational modes. We see that in every case where bands don't mix, i.e. the bottom three bands, the $E$ bands slope downwards in frequency from $\Gamma$ to $X$, while the $O$ band slopes upwards.  In plots of the mechanical modes, color indicates the magnitude of the displacement field (blue no displacement, red large displacement) and the displacement of the structure has been exaggerated for viewing purposes.}
\label{fig:tight_bindings}
\end{center}
\end{figure}

As will be evident below in the design of phononic waveguides and cavities, it is useful to study the properties of the tight-banding bands in a little more detail.  Considering the simple linear lattice example introduced above (refer to Fig.~\ref{fig:1dcrystal}(b)), the group of the wavevector at the $\Gamma$ and $X$ points of the Brillouin zone possesses the full point group symmetry of the crystal itself.  As such, the Bloch modes ($\m Q$) at these high symmetry points can be characterized according to their vector symmetry with respect to reflection $\sigma_x (x,y,z) = (-x, y,z)$ about the $x$-axis in a plane intersecting the middle of the unit cell of the linear nanomechanical structure.  We have for $x$-symmetric ($\m Q^{(x+)}$) and $x$-antisymmetric ($\m Q^{(x-)}$) Bloch modes, for which $\sigma_x \m Q^{(x\pm)} (\sigma_x \m r) = \pm \m Q^{(x\pm)}(\m r)$, the following relation for the displacement vector ($\m Q$) at the unit-cell boundaries ($\m r_b$):
\begin{eqnarray}
\m Q^{(x+)} (\m r_b +a \m e_x) &=& + \sigma_x \m Q^{(x+)} (\m r_b ), \\
\m Q^{(x-)} (\m r_b +a \m e_x) &=&  -\sigma_x \m Q^{(x-)} (\m r_b ), 
\end{eqnarray}
On the other hand, we have the usual phase shift acquired by the different Bloch modes, which for the $X$-point and $\Gamma$-point modes yields,
\begin{eqnarray}
\m Q_\Gamma(\m r + a\m e_x) &=& + \m Q_\Gamma(\m r), \\
\m Q_X(\m r + a\m e_x) &=& - \m Q_X(\m r).
\end{eqnarray}
These constraints together imply that $x$-symmetric modes at the $\Gamma$- and $X$-points must obey respectively the conditions 
\begin{eqnarray}
\m e_x \cdot \m Q^{(x+)}_\Gamma(\m r_b)  &=& 0,\label{eqn:QxpGamma}\\
\m e_x \times \m Q^{(x+)}_X(\m r_b)   &=& \m 0\label{eqn:QxpX}.
\end{eqnarray}

In many cases the excitation is of a dominant polarization. For example, for the $(\sigma_y,\sigma_z) = (+,+)$ vector symmetry vibrational modes, the dominant polarization is found to be $Q_x$ (not suprising as for this symmetry the displacement in $y$ or $z$ would have to be of higher order, involving a stretching/compression of the membrane along these directions). For this symmetry of modes then, the boundary condition given by Eqn. (\ref{eqn:QxpGamma}) is approximately equivalent to a fixed boundary condition, whereas the boundary condition of Eqn. (\ref{eqn:QxpX}) can be approximated as a free boundary condition. Since intuitively, one expects the frequency of a free resonator to increase by fixing part of it, the above argument implies that for the given polarization  mode symmetry, $(\sigma_y,\sigma_z) = (+,+)$, $x$-symmetric modes bend up while $x$-antisymmetric modes slope downwards as the in-plane wavevector $k$ varies from $\Gamma-X$. This correspondence is elucidated in Figure \ref{fig:tight_bindings}, where the tight-binding bands are shown for the aforementioned quasi-1D system. In all cases, the frequencies calculated with fixed or free boundary conditions on the square drumhead correspond to their respective high symmetry Bloch function frequencies to within a factor $10^{-2}$. In what follows, we use these features of the tight-binding bands to help identify vibrational bands that will strongly couple to optical waves.  

\begin{figure}[t]
\begin{center}
\includegraphics[width=0.80\columnwidth]{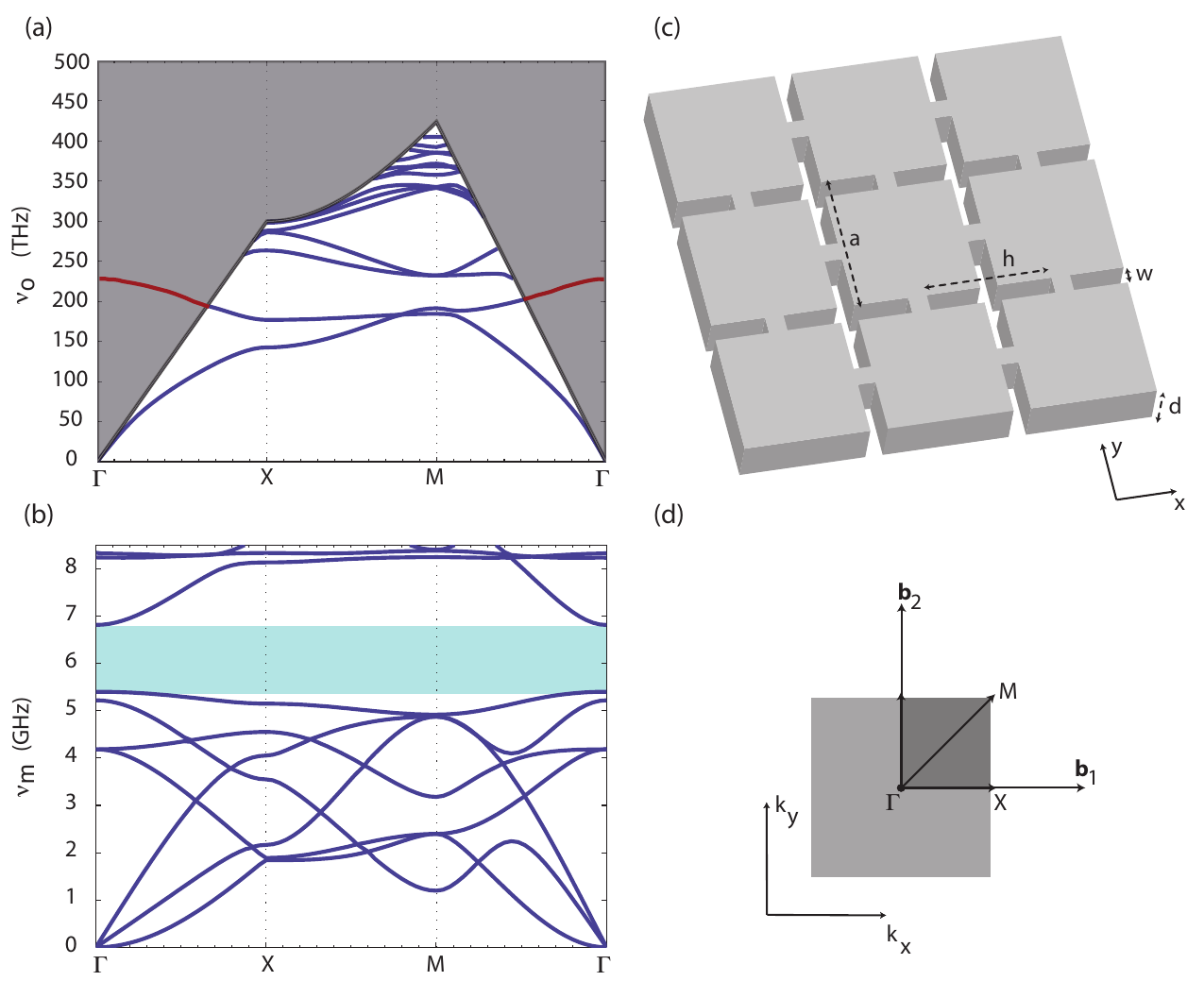}
\caption{(a) Photonic and (b) phononic bandstructure of the quasi-2D Si cross substrate shown schematically in (c) real space and (d) reciprocal space.  The photonic bandstructure is for the even symmetry modes of the slab only.  The band diagrams were calculated for a Si structure with parameters $(d,h,w,a)=(220,200,100,500)~\text{nm}$. Notably, there is not even a photonic pseudo in-plane bandgap for the even vertical symmetry optical modes of the slab.}
\label{fig:2dcrystal_cross}
\end{center}
\end{figure}

\subsection{Quasi-2D ``Cross'' Crystal}

\begin{figure}[t]
\begin{center}
\includegraphics[width=0.80\columnwidth]{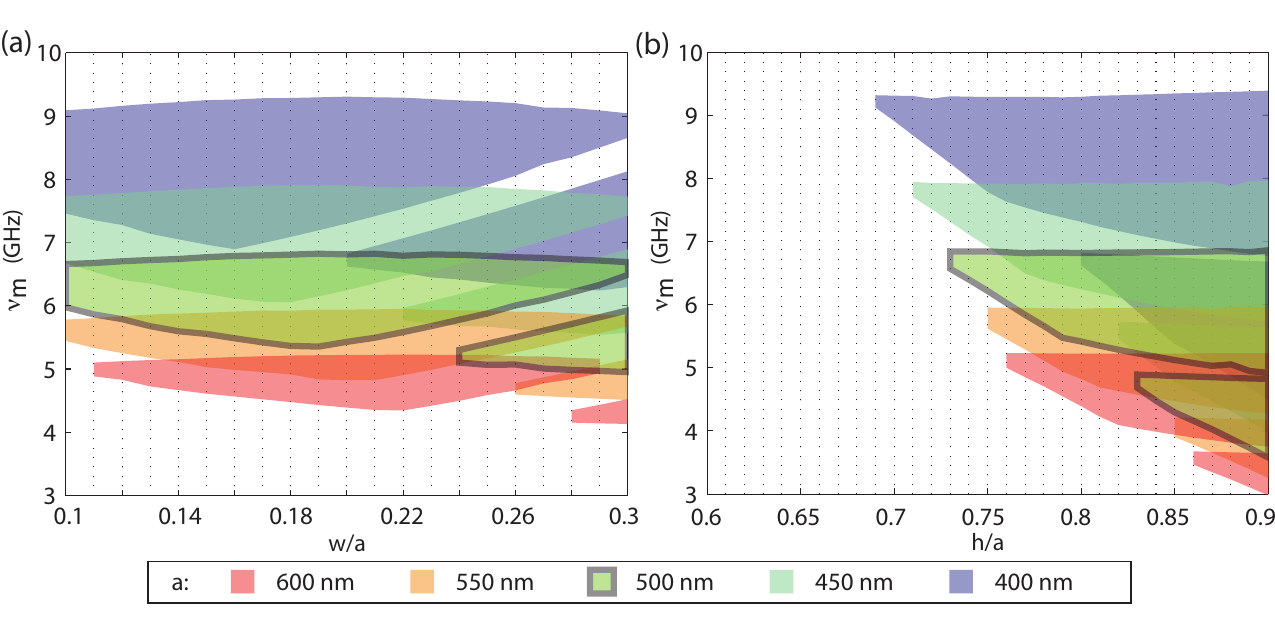}
\caption{Phononic gap maps for a Si quasi-2D cross structure. The cross width is varied in (a) and the cross height is varied in (b). The nominal structure about which the variations are performed is characterized by parameters $(d,h,w,a)=(220,200,100,500)~\text{nm}$.  The lattice constant $a$ is varied while keeping the ratios $h/a = 0.8$ and $w/a=0.2$ constant in (a) and (b), respectively.}
\label{fig:phononic_gapmap_cross}
\end{center}
\end{figure}

In going to a quasi-2D system, we begin with the simplest extension of the quasi-1D structure of Fig.~\ref{fig:1dcrystal}(b), that being the same elements of square drumheads and thin connectors, but now arrayed in two dimensions as shown in Fig.~\ref{fig:2dcrystal_cross}(c). We call this the ``cross'' substrate, since it results from a square array of crosses cut into a slab. Each cross is characterized by a height $h$ and a width $w$, which along with the lattice spacing $a$ and slab thickness $d$, serve to fully parametrize geometrically the system.  For reference, the reciprocal space representation of the lattice is shown in Fig.~\ref{fig:2dcrystal_cross}(d), in which the common notation of the high symmetry points of the first Brillouin zone in a square lattice are used. The phononic bandstructure, including all symmetries of vibrational modes, of the cross substrate is shown in Fig.~\ref{fig:2dcrystal_cross}(b) for the same set of parameters as used in the quasi-1D structure of Fig.~\ref{fig:1dcrystal}.  Clearly this Si structure has excellent mechanical properties, with a large bandgap opening up between $5.3$ and $6.8$ GHz.  The phononic gap maps for a variety of relevant parameters of the cross substrate are shown in Figure~\ref{fig:phononic_gapmap_cross}, indicating the robustness of the phonon bandgap to each parameter.  

Despite the encouraging  mechanical properties of the cross substrate, as we will see it is difficult to realize this substrate as an OMC due to its unfavorable optical properties.  In general, design of photonic bandgaps in materials with refractive index of order $n \sim 3$ (semiconductors) is more suited to a plane-wave expansion approach as opposed to the tight-binding picture discussed above for nanomechanical vibrations.  The square lattice, with its low symmetry, behaves differently for plane waves propagating in different directions, such as at the high symmetry $X$ and $M$ points of the first Brillouin zone boundary.  This results in a much smaller in-plane photonic bandgap for the square lattice in comparison to a higher symmetry lattice such as the hexagonal lattice.  Note also that for quasi-2D photonic structures we usually talk about an \emph{in-plane} bandgap only, as light (unlike sound) can also propagate vertically into the surrounding vacuum cladding.  The photonic bandstructure of the Si cross substrate, assuming a refractive index for Si of $3.4$, is shown in Fig.~\ref{fig:2dcrystal_cross}(a) for the even vertically symmetric optical modes of the Si slab (these modes include the fundamental TE-like bands).  These and the following photonic simulations were done using the software package MPB \cite{Johnson2001:mpb}.  The photonic bandstructure consists of two distinct regions: (i) the guided mode region below the light line (shown as a black line) in which there are a discrete number of mode bands, all of which are evanescent into the surrounding cladding, and (ii) the region above the light (shown in grey) in which \emph{leaky} guided mode resonances and a continuum of radiation modes exist.  We have shown the extension of the lowest lying guided mode band above the light line as a red line, where it behaves as a \emph{leaky}, but highly localized resonance of the Si slab.  

\begin{figure}[t]
\begin{center}
\includegraphics[width=0.80\columnwidth]{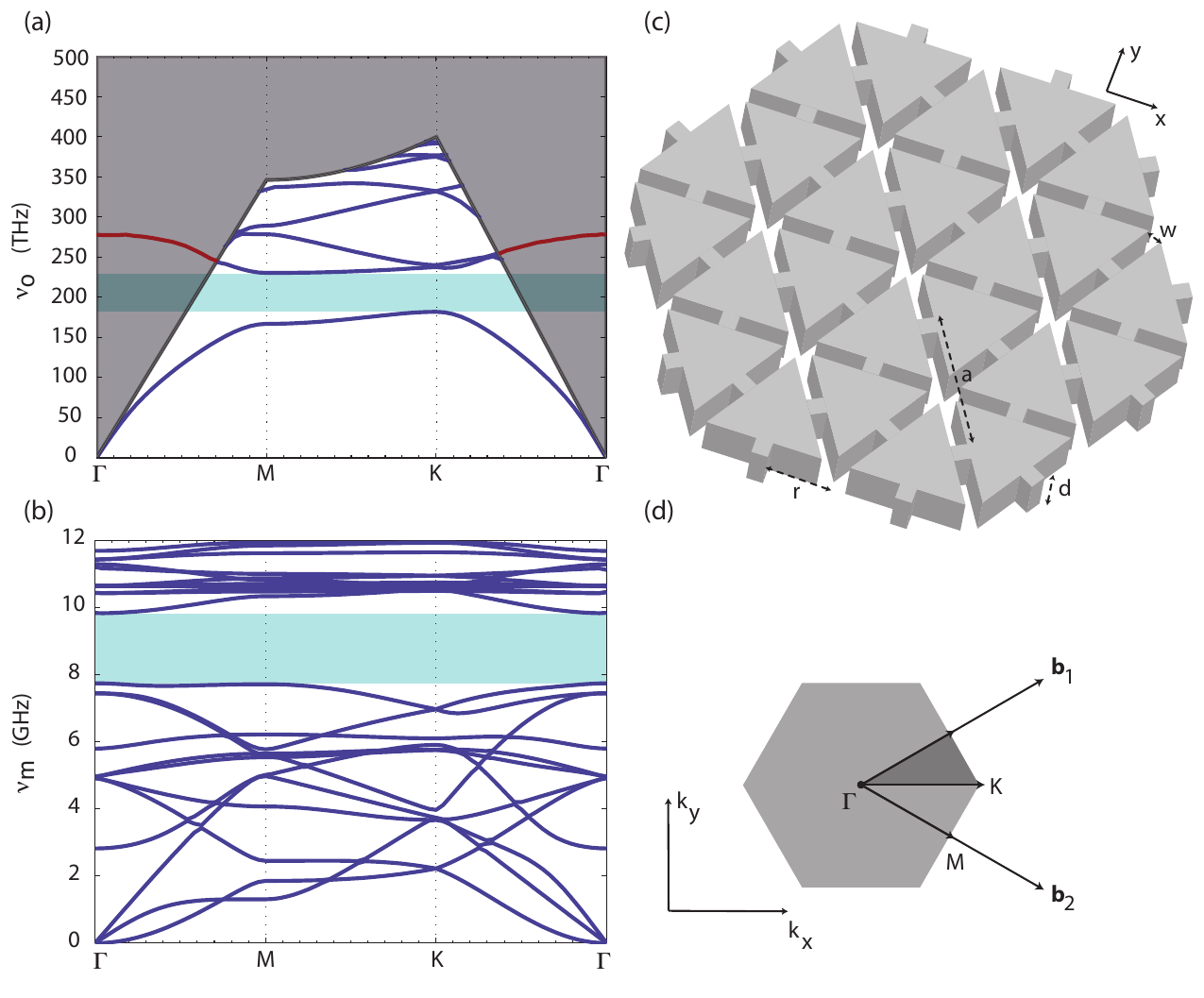}
\caption{(a) Photonic and (b) phononic bandstructure of the quasi-2D Si snowflake substrate.  Schematic of the snowflake slab substrate in (c) real space and (d) reciprocal space.  The band diagrams were calculated for a Si structure with parameters $(d,r,w,a)=(220,200,75,500)~\text{nm}$. For these parameters, there are large simultaneous phononic and photonic bandgaps.}
\label{fig:2dcrystal_snowflake}
\end{center}
\end{figure}

Due to the presence of the continuum of radiation modes above the light line, it is clear that in the photonic case one can only talk about a pseudo in-plane bandgap in the case of a quasi-2D slab structure.  Much more problematic is the presence of \emph{leaky} guided mode resonances above the light line, which can strongly couple to the guided modes of the structure in the presence of perturbations of the lattice.  These perturbations can both be unintentional, such as in fabrication imperfections, or intentianal such as in the formation of resonant cavities and linear waveguides.  Here we will adopt a practical definition of a pseudo in-plane photonic bandgap as one where the bandgap extends across \emph{both} guided mode and leaky resonances.   As can be seen in the photonic bandstructure of Fig.~\ref{fig:2dcrystal_cross}(d), the cross structure lacks even a pseudo in-plane photonic bandgap for the TE-like modes of the slab structure (the odd symmetry modes of the slab, which include the fundamental TM-like modes, do not even possess a guided mode bandgap).  Due then, to leaky resonances with large local density of states in the slab, the cross substrate is ruled out as a suitable structure from which to form important photonic elements such as ultrahigh-$Q$ optical cavities. This is the same problem faced by the ``honey-comb'' crystal proposed in Ref. \cite{Mohammadi2008}.

\subsection{Quasi-2D ``Snowflake'' Crystal}\label{ss:snowflake}

\begin{figure}[t]
\begin{center}
\includegraphics[width=0.85\columnwidth]{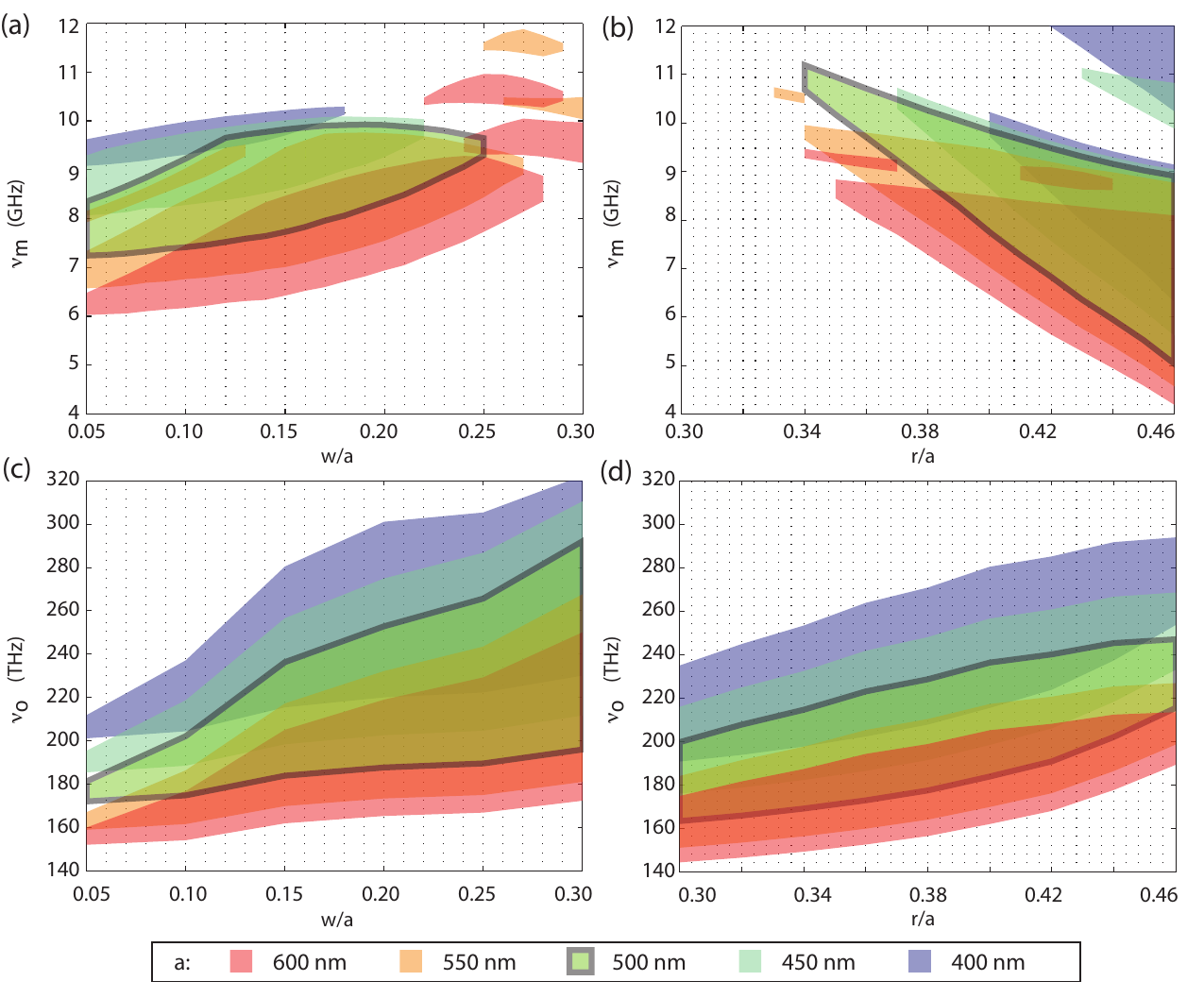}
\caption{Phononic (a,b) and photonic (c,d) gap maps for a quasi-2D snowflake substrate. The photonic gap maps are for the even vertical symmetry modes of the slab only.  In (a,c) we vary the snowflake width and (b,d) the snowflake radius. The nominal structure about which the variations are performed is characterized by parameters $(d,r,w,a)=(220,200,75,500)~\text{nm}$. The lattice constant $a$ is varied while keeping the ratios $r/a = 0.4$ and $w/a=0.15$ constant in (a,c) and (b,d), respectively.}
\label{fig:gapmap_snowflake}
\end{center}
\end{figure}

The hexagonal lattice counterpart of the cross substrate is shown in Fig.~\ref{fig:2dcrystal_snowflake}, which we term the ``snowflake'' substrate. Each snowflake pattern is characterized by radius $r$, and a width $w$; which along with the lattice spacing $a$ and slab thickness $d$, serve to fully parametrize geometrically the system. The phononic gap maps for a variety of relevant parameters of the snowflake substrate are shown in Figs.~\ref{fig:gapmap_snowflake}(a) and ~\ref{fig:gapmap_snowflake}(b). The snowflake substrate, unlike the cross substrate, does possess favorable optical properties. Gap maps for the photonic properties are shown in Figs.~\ref{fig:gapmap_snowflake}(c) and ~\ref{fig:gapmap_snowflake}(d), where again we focus on the even vertical symmetry optical modes of the slab (which includes the fundamental TE-like bands of primary interest).  Due to the fact that the phononics is more sensitive to connector width, $a-2r$, while the photonics is more sensitive to the air-slot width, $w$, this crystal provides us with two different tunable parameters for control over the photonic and phononic properties of the system. This property, which is apparent from the respective gap maps, is highly valuable as it pertains to designing optomechanical devices which simultaneous must manipulate sound and light.  For this reason, and for its superior optical bandgap properties, in what follows we focus on the snowflake crystal.  We begin with a study of the light and sound waveguiding properties of this structure.  

\definecolor{WGeyez}{rgb}{0.1,0.1,0.9}
\definecolor{WGoz}{rgb}{0.7,0.9,0.7}
\definecolor{WGoyez}{rgb}{0.9,0.7,0.7}
\definecolor{WGleaky}{rgb}{0.7,0.3,0.3}

\begin{figure}[t]
\begin{center}
\includegraphics[width=0.9\columnwidth]{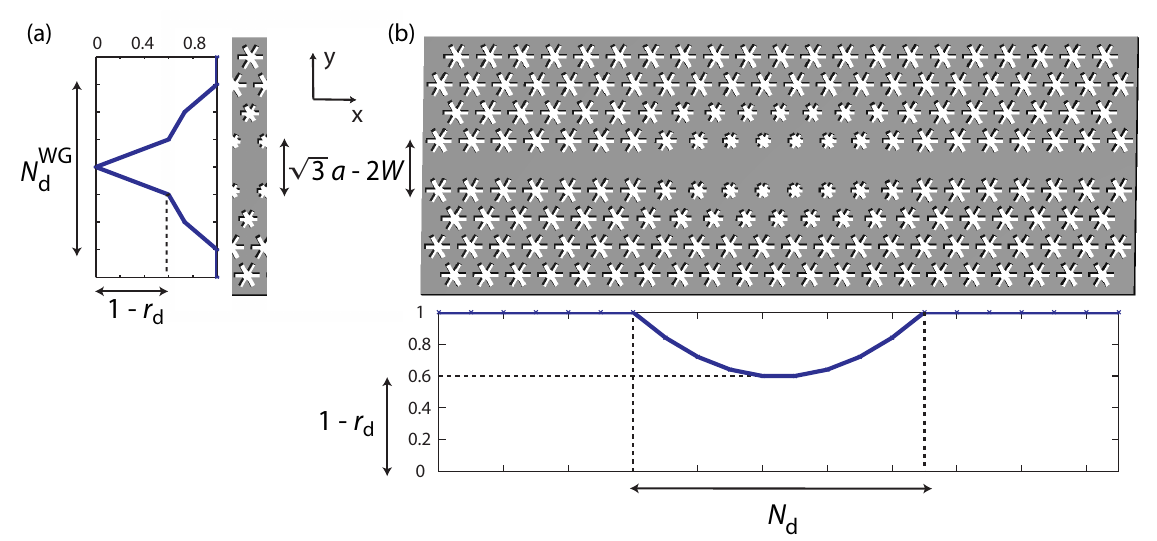}
\caption{(a) Schematic of a $W1$-like linear defect waveguide in the quasi-2D snowflake crystal slab structure.  The central row of snowflake holes is completely removed along the $x$-direction ($\Gamma$-$K$ in reciprocal space), and the remaining top and bottom pieces of the lattice are shifted by a value $W$ towards each other (effectively a strip of $2W$ is removed from the center of the waveguide).  A transverse radius variation of the snowflake holes is also applied.  $N_d^{\text{WG}}$ is the number of rows of holes which take part in forming this defect.  The  number $r_d$ represents the factor by which the radius of holes on the two rows neighbouring the center of the defect are reduced; i.e. the radius is changed to $r\times(1-r_d)$, where $r$ is the nominal radius. Rows going further out from the center of the waveguide have radii which scale quadratically to the nominal value of $r$. (b) Cavities are formed from this line-defect waveguide by a longitudinal modulation of the waveguide parameters.  In this case, the $r_d$ scale factor is varied quadratically from $0$ to a desired value at the cavity center along the length of the waveguide over a period of $N_d$ lattice periods. The cavity structure shown here has $r_{d} = 0.4$ at the cavity center, $N_d = 10$ and $N_d^{\text{WG}} = 7$.}
\label{fig:dim_diagram}
\end{center}
\end{figure}

\section{Waveguide Design}\label{sec:waveguide}

\begin{figure}[t]
\begin{center}
\includegraphics[width=0.75\columnwidth]{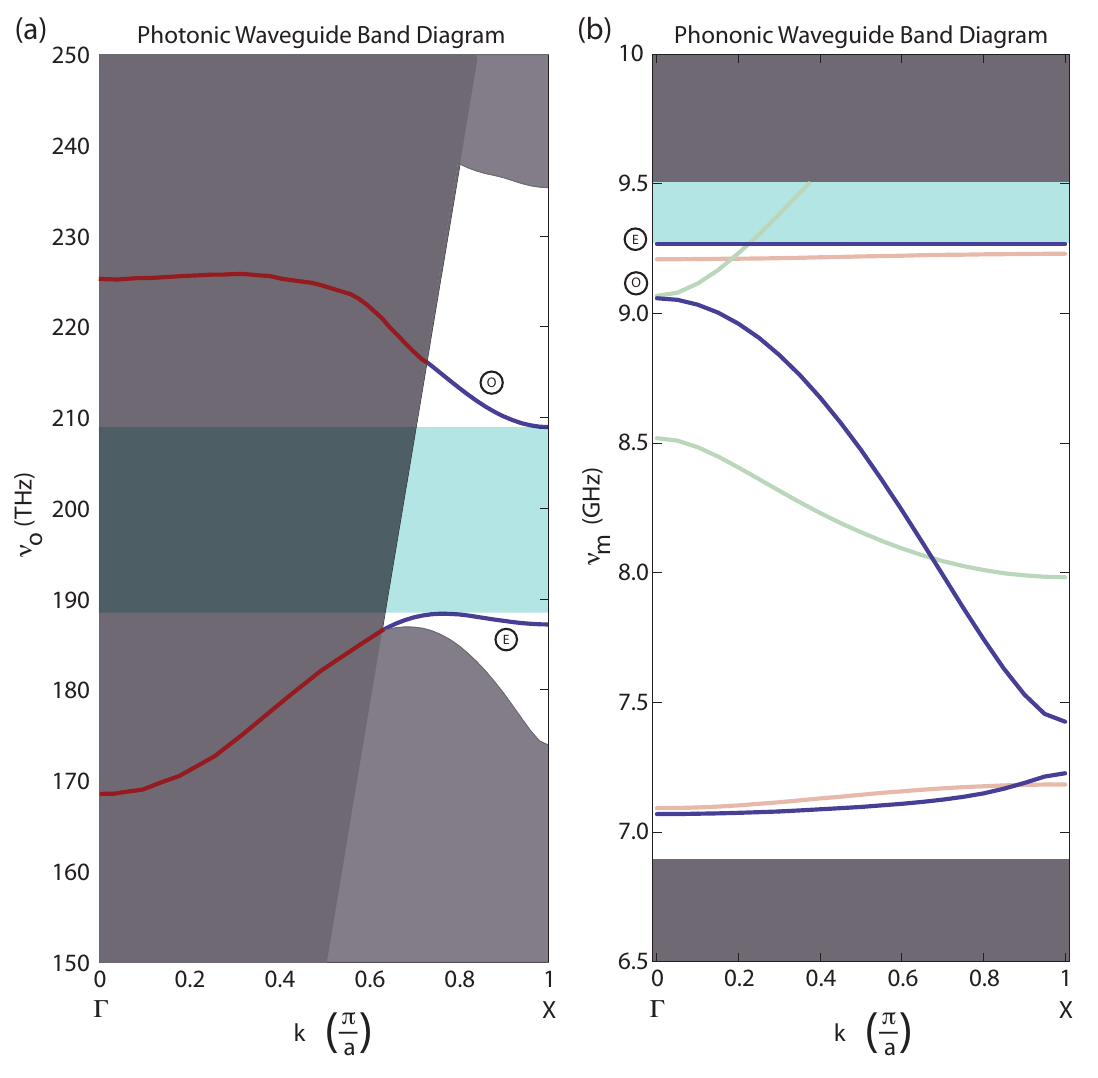}
\caption{(a) Photonic and (b) phononic waveguide bands for an optomechanical waveguide on a snowflake substrate with $(d,r,w,a)=(220,210,75,500)~\text{nm}$, and a $W1$-like waveguide with properties $W = 200~\text{nm}$ and $r_d = 0$. The $E$ and $O$ symbols represent even and odd vector parity, respectively, of the zone boundary modes with respect to $\sigma_y$ reflections for the optical and $\sigma_x$ reflections for the mechanical modes.  The same mode labels are used in Figure \ref{fig:waveguide_tuning}. In (a), only the even vertical symmetry modes are shown (those including the fundamental TE-like modes, but not the fundamental TM-like modes).  In (b), guided modes with different transverse symmetries $(\sigma_y,\sigma_z)$ are colored ({\color{WGeyez}---}), ({\color{WGoz}---}) and ({\color{WGoyez}---}) for symmetries $(+,+)$, $(\pm,-)$, and $(-,+)$, respectively.  In both diagrams frequencies above and below the in-plane bandgap are colored light grey, and in (a) the light cone is region is colored dark grey. Above the light-line in (a), the leaky mode bands are colored red ({\color{WGleaky}---}). In both (a) and (b) we have highlighted the bandgap regions which will be relevant in the cavity design pursued below.}
\label{fig:waveguide_bands}
\end{center}
\end{figure}

A line-defect in a photonic bandgap material will act as a waveguide for light~\cite{ref:Chutinan2000,ref:Johnson4}. In a quasi-2D crystal structure with simultaneous in-plane photonic and phononic bandgaps, such a defect should be able to direct both light and sound around in the plane of the crystal.  In this section, we study the guided modes in these types of linear-defect waveguides of the snowflake crystal slab. It is of interest to understand the properties of these modes, since it has been shown for photonic crystal slabs that good waveguides also yield ultrahigh-$Q$ optical cavities~\cite{Song2005,Kuramochi2006}. In Section \ref{sec:cavity} we follow this design technique to demonstrate ultrahigh-$Q$ optomechanical cavities in which both light and sound are effectively localized in the same volume with very little radiative loss. Although we are primarily interested here in \textit{cavity} optomechanics, the coupling between guided photons and phonons in periodic structures is also an interesting subject of study which has been recently explored in photonic crystal fiber systems~\cite{Kang2009}. As will be discussed further below, studying \textit{guided-mode} optomechanical properties also allows one to simplify the design and optimization of  \textit{cavity} optomechanical devices.  

When designing optical and mechanical cavities and waveguides, it is desirable to have control over where the waveguide bands are placed within the frequency bandgap~\cite{ref:Chutinan2000}.  For example, previous demonstrations of photonic crystal cavities have often involved a line defect waveguide in which a localized cavity resonance was formed by locally tuning the line defect guided mode out of the bandwidth of the waveguide band and into the bandgap. In previous work, this tuning has been achieved by changing locally the longitudinal lattice constant along the guiding direction~\cite{Song2005}, the width of the line-defect forming the waveguide~\cite{Kuramochi2006}, or the radii of the holes adjacent to the line-defect region~\cite{Barclay2003}.   Figure~\ref{fig:dim_diagram}(a) shows an example of a linear defect waveguide formed in the snowflake crystal slab.  This waveguide consists of a row of snowflake holes which have been removed (a \emph{W1}-like waveguide), and a transverse variation of the slowflake hole size has been applied (see Fig.~\ref{fig:dim_diagram} caption for details).  Figure~\ref{fig:dim_diagram}(b) shows a corresponding resonant cavity structure formed from the linear defect waveguide.   

The photonic and phononic bandstructures of a linear defect waveguide with $W = 200~\text{nm}$ is shown in Fig. \ref{fig:waveguide_bands}.  In this diagram only the vertically ($z$) symmetric optical bands are shown.  The waveguide dielectric structure also has a transverse mirror symmetry, $\sigma_y$, about the $y$-axis in the middle of the waveguide.  The transverse symmetry of each of the mechanical bands is indicated in Fig.~\ref{fig:waveguide_bands} by the color of the band, whereas for the optical waveguide bands we use the labels $E$ (even) and $O$ (odd) to indicate the $\sigma_{y}$ parity of the fields.  A similar $E$ and $O$ labelling scheme is used for the mechanical waveguide modes at the $\Gamma$-point, although in this case the parity relates to the $\sigma_{x}$ symmetry of the mechanical displacement field within each unit cell along the $x$-direction of the waveguide.  Also, in the photonic band diagram we have indicated the light cone with a dark grey shade.  The regions above and below the guided mode bandgap of the unperturbed snowflake crystal are shaded a light grey in both the photonic and phononic band diagrams, with the leaky regions of the photonic waveguide  bands colored in red.  For this waveguide width, a significant pseudo-bandgap can be seen in the photonic bandstructure (shaded in light blue).  At the same time, several phononic bandgaps can be seen in the bands of the mechanical band diagram of the waveguide. Our primary interest when forming a resonant cavity in the next section will be the bandgap highlighted in light blue between the highest frequency phononic waveguide band and the upper frequency band-edge of the unperturbed snowflake crystal.  This mechanical waveguide band has the desirable property of very flat dispersion which allows for highly localized phonon cavity states in the presence of waveguide perturbations.  

\begin{figure}[t]
\begin{center}
\includegraphics[width=0.9\columnwidth]{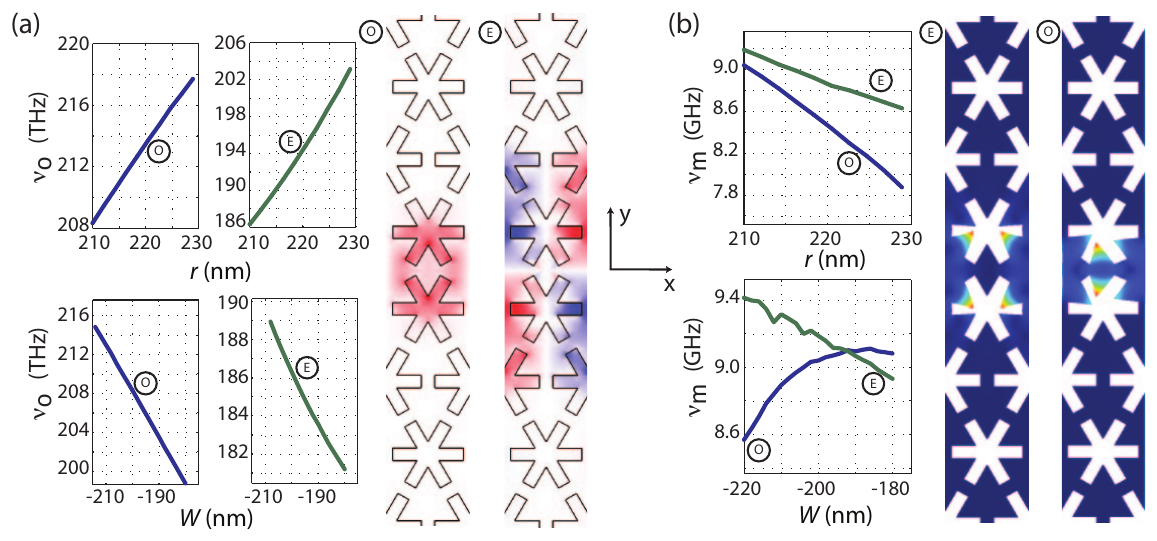}
\caption{Tuning of the (a) $X$-point optical and (b) $\Gamma$-point mechanical waveguide modes of a $W1$-like waveguide in a snowflake substrate with parameters $(d,r,w,a)=(220,210,75,500)~\text{nm}$.  Tuning is shown versus both waveguide width $W$ and snowflake radius $r$. In this case we have taken $r_d = 0$, and $N^{\text{WG}}_d = 1$.  See Fig.~\ref{fig:dim_diagram}(a) for line-defect waveguide description. The $E$ and $O$ symbols represent respectively the even and odd vector parity of $\m E (\m r)$  with respect to mirror reflection $\sigma_y$ about the middle of the waveguide and $\m Q(\m r)$ with respect to mirror reflection $\sigma_x$ about the middle of each waveguide unit cell.  In the optical field plots we show a snapshot in time of the $y$-polarization of the electric field ($E_y(\m r)$), with red and blue indicating positive and negative values of the field, respectively.  In the mechanical mode plots, color indicates the magnitude of the displacement field (blue no displacement, and red large displacement), and the displacement of the structure has been exaggerated for viewing purposes.}
\label{fig:waveguide_tuning}
\end{center}
\end{figure}

In the design of an optomechanical device, one in which light and sound must be simultaneously manipulated, the independent control of the two types of wave excitations is desired.  The tuning of the optical and mechanical waveguide bands of a $W1$-like line-defect waveguide is shown in Fig.~\ref{fig:waveguide_tuning} for two different types of waveguide geometry perturbations.  For simplicity we have only shown the tuning of the waveguide modes at the zone boundary ($X$-point for the optical and $\Gamma$-point for the mechanical waveguide modes).  From these plots, it is evident that radius modulations of the snowflake hole tend to tune the optical and mechanical modes in differing directions, whereas for width modulations (through $W$) of the waveguide the optical and mechanical frequencies tend to tune in a similar direction (this is not true of the odd symmetry mechanical mode in this narrow waveguide). A heuristic argument for this behavior goes as follows.  For an optical mode, regions of high refractive index, such as the silicon, tend to reduce the optical frequency for a given curvature (wavevector) of the optical wave.  Quite the opposite is true for mechanical excitations in which the material adds to the stiffness of the structure, thereby generally raising the frequency of acoustic waves. This suggests that by using a perturbation where the hole sizes are slightly reduced, since we are increasing the amount of high-index(stiffness) material the photon (phonon) sees while keeping the wavelength constant, the frequency of the mode will decrease (increase). On the other hand, when the waveguide width is increased, one is in some sense increasing the ``transverse'' lattice constant of the crystal. Since the lattice constant sets the wavelength for both the optical and mechanical modes, increasing it will cause the frequencies of both waves to drop. By using these perturbations simultaneously, within a certain small range, photonic and phononic bands may be raised and lowered independently. This is a powerful consequence of using differing tuning mechanisms, and allows us to design independently the longitudinal (cavity) confining potentials for phonons and photons.

\section{Optomechanical Coupling Relations}\label{ss:WG_coupling}

In the design methodology followed in this paper, a resonant cavity is formed by locally modulating the properties of a linear defect, or waveguide, in a planar crystal structure.  This methodology has been used previously in the design of high-$Q$ photonic crystal cavities, and due to the similarities in the kinematic properties of the wave equations of phonons and photons, we expect it to also produce phononic crystal cavities in the snowflake crystal structure. For our purposes, however, having the photonic and phononic resonances simply co-localized doesn't suffice, as their interaction must also be tailored, and maximized. To better understand the origin of the optomechanical coupling, it is useful to study the interaction at the level of the waveguide modes, from which the localized cavity resonances are formed.

The coupling between guided optical and mechanical waves has been studied previously in both theoretical and experimental settings \cite{Dainese2006,Kang2009} for photonic crystal fiber structures with \textit{continuous} longitudinal symmetry. These analyses have generally expanded on calculations of acousto-optical scattering in bulk materials~\cite{Boyd2008}. For this work, we are interested in the case of \textit{discrete} longitudinal symmetry, and a calculation of the coupling per unit cell. Instead of extending the aforementioned analysis to the case of discrete longitudinal symmetry, our approach will be to start with the known \emph{cavity} optomechanical coupling relations, and then to work backwards to a relavant per unit cell guided-mode coupling.  This has the benefit of providing a direct relation between the guided-mode and cavity-mode optomechanical couplings.  Specifically, using Johnson's formulation of perturbation theory for moving dielectric boundaries \cite{Johnson2002}, which has been previously applied successfully to the calculation of cavity optomechanical properties \cite{Eichenfield2009d,Eichenfield2009c}, we find the cavity optomechanical coupling in terms of the localized mechanical vibration field and an effective optical energy density. Then, using the Wannier function formalism \cite{Wannier1962,Charbonneau-Lefort2002,Istrate2002,Painter2003}, we relate (approximately) the cavity and waveguide modes to one another, providing a relation for the guided-mode optomechanical coupling from the cavity-mode optomechanical coupling. 

The formula for the (lowest-order) optomechanical coupling rate in a deformable cavity has been shown to be most generally given by \cite{Eichenfield2009c}:
\begin{equation}
g = \sqrt{\frac{\hbar}{2 \Omega}}\frac{\omega_o}{2}\frac{\int \left(\m Q(\m r) \cdot \m n\right)(\Delta \epsilon |\m E^\parallel |^2 - \Delta(\epsilon^{-1})|\m D^\perp |^2) \text{d}A}{\sqrt{\int \rho  |\m Q(\m r)|^2 \text{d}^3\m r} \int \epsilon(\m r)|\m E(\m r)|^2 \text{d}^3\m r}
\end{equation}
This is a pure rate, and is found by multiplying the dispersion of the optical cavity resonance with mechanical oscillator displacement ($g_\text{OM}\equiv \partial\omega_{c}/\partial x$) by the zero-point fluctuation amplitude of the mechanical oscillator ($x_\text{ZPF} = \sqrt{\frac{\hbar}{2 m_\text{eff}\Omega}}$). To relate the cavity optomechanical coupling to the properties of the waveguides, we assume that our acoustic and optical cavity fields can both be written in terms of a waveguide Bloch function multiplied by a smoothly varying envelope function. In general these cavity fields can be represented as superpositions of terms of the type $\m E_\pm (\m r) = \m E_{e}(\m r) e^{\pm i\m k_e \cdot \m r} f_e(x)$ ($\m Q_\pm (\m r) = \m Q_{m}(\m r) e^{\pm i\m k_m \cdot \m r} f_m(x)$), where $\m E_{e}$ ($\m Q_{m}$) is a periodic Bloch function, $k_e$ ($k_m$) the reduced wavevector, and $f_e(x)$ ($f_m(x)$) the envelope of the electric (mechanical displacement) cavity field.  Note that both the co- and counter-propagating terms ($\pm k_{e,m}$) are necessary to describe the localized standing-wave resonances of a linear cavity. 

While a general analysis is possible, we limit ourselves here to the case where the optical cavity mode is formed from the $X$-point of the waveguide band diagram (see Fig.~\ref{fig:waveguide_bands}(a)), with  $\m k_e = \m k_X$, and the mechanical cavity mode is a $\Gamma$-point mode with $\m k_m = 0$. The condition on the optical mode is necessary in a quasi-2D slab structure to achieve a high-$Q$ optical cavity, as small $k$-vector components in the plane of slab can radiate into the light cone of the low-index cladding surrounding the slab.  The mechanical mode condition is a phase matching requirement for the coupling of the two counter-propagating optical waves of a standing-wave cavity resonance, $\m k_X + (-\m k _X) = \m k_m = 0$. Hence, starting with

\begin{eqnarray}
\m E_\pm (\m r) &=& \m E_{X}(\m r) e^{\pm i\m k_X \cdot \m r} f_e(\m x),\\
\m Q (\m r) &=& \m Q_{\Gamma}(\m r) f_m(\m x),
\end{eqnarray} 

\noindent and assuming that the envelope functions vary slowly over a lattice spacing, and that they have no zero-crossings, we separate the integrals into a product of two integrals, one over a single waveguide unit-cell and the other across multiple unit-cells. For example,

\begin{equation}
\int |\m Q(\m r)|^2 \text{d}\m r \approx \frac{1}{a} \int_\Delta |\m Q_{\Gamma}(\m r)|^2 \text{d}^3 \m r \int |f_m(x)|^2 dx. \label{eqn:separability}
\end{equation}

\noindent From here, we arrive at the following expression for $g$:

\begin{equation}
g \approx \sqrt{a} g_\Delta \frac{\lt f_e | f_m  | f_e \rt}{\sqrt{\lt f_m | f_m \rt} \lt f_e | f_e \rt}, \label{eqn:g_and_gDelta}
\end{equation}

\noindent where $g_\Delta$ is the guided-mode optomechanical coupling given by

\begin{equation}
g_\Delta =  \sqrt{\frac{\hbar}{2 \Omega}}\frac{\omega_o}{2}\frac{\int_\Delta \left(\m Q_{\Gamma}(\m r) \cdot \m n\right)(\Delta \epsilon |\m E^\parallel_{X} |^2 - \Delta(\epsilon^{-1})|\m D^\perp_{X} |^2) \text{d}A}{\sqrt{\int_\Delta \rho |\m Q_{\Gamma}(\m r)|^2 \text{d}^3\m r} \int_\Delta \epsilon(\m r)|\m E_{X}(\m r)|^2 \text{d}^3\m r}. \label{eqn:gDelta}
\end{equation}

Equation (\ref{eqn:g_and_gDelta}) shows that the optomechanical coupling achievable is the product of a term $g_\Delta$ depending only on the linear waveguide properties and a second term which is a function of the envelope functions $f_e(x)$ and $f_m(x)$ describing the localization of the cavity resonances along the length of the waveguide. In many relevant systems, the optical and mechanical modes may be approximated as Gaussians with standard deviation in intensity profile of $L_m$ and $L_e$.
In this case the envelope-dependent component of the cavity optomechanical coupling is,

\begin{eqnarray}
\lt f_e | f_m | f_e \rt =  \frac{1}{\left( 2 \pi \right)^{\frac{1}{4}}} \frac{1}{\sqrt{L_m + \frac{1}{2}\frac{L_e^2}{L_m}}}. \label{eqn:omc_inner_p}
\end{eqnarray}

\noindent The largest value of the envelope dependent part of the optomechanical coupling from equation (\ref{eqn:omc_inner_p}) is achieved by making $L_m = L_e/\sqrt{2}$.  For this ratio of mechanical and optical cavity Gaussian profiles one arrives at a maximum optomechanical coupling rate of
 
\begin{eqnarray}
g_\text{optimal} \approx \frac{1}{\left( 4 \pi \right)^{\frac{1}{4}}} g_\Delta \sqrt{\frac{a}{L_e}}. \label{eqn:g_optimal}
\end{eqnarray}

\noindent Clearly, the more localized the optical and mechanical resonances the larger the optomechanical coupling, all other things being equal. 

\section{Optomechanical Cavity Design}\label{sec:cavity}

As mentioned above, higher-$Q$ optical cavity resonances will be formed from optical modes near the $X$-point of the linear-defect waveguide as they lie underneath the out-of-plane light cone.  Phase-matching then requires the mechanical resonance to be formed primarily from the $\Gamma$-point in order to have significant optomechanical interaction of the two localized resonances.  Given the even symmetry along $x$ within each unit cell of the \emph{intensity} of the optical field for waveguide modes at the $X$-point (again, all $X$-point modes can be classified by their $\sigma_{x}$ parity, and thus their intensity must be symmetric), and considering the form of Eqn. (\ref{eqn:gDelta}) for the per unit cell optomechanical coupling, we see that only the even symmetry ($E$) mechanical modes at the $\Gamma$-point yield a non-zero $g_{\Delta}$.  As such, we choose to form the localized phononic cavity resonance from the uppermost phononic waveguide band in Figure \ref{fig:waveguide_bands}(b) (the lowermost phononic waveguide band, also of even parity at the $\Gamma$-point, was found to have a smaller $g_{\Delta}$). To avoid coupling to mechanical waveguide bands below this upper band, we choose to form a cavity defect perturbation which \textit{increases} the frequency of the upper mechanical waveguide band, thus localizing the phononic resonance in the highlighted blue pseudo-bandgap of Fig.~\ref{fig:waveguide_bands}(b). 

In the case of the optical field, we can choose to form the localized cavity resonance from either the upper or the lower frequency waveguide bands that define the photonic pseudo-bandgap of the waveguide (see Fig.~\ref{fig:waveguide_bands}(a)).  We choose here to use the upper frequency waveguide band due to its more central location in the pseudo-bandgap of the unperturbed snowflake crystal.  The curvature of the upper frequency waveguide band near the $X$-point is positive, so we need to have the cavity perturbation cause a local \textit{decrease} in the band-edge frequency; the opposite frequency shift required for that of the mechanical waveguide band. From Figure \ref{fig:waveguide_tuning}, it is evident that the radius modulation satisfies this requirement, i.e. it tunes the optical and mechanical band-edge modes in opposing directions. Here we use a combination of transverse and longitudinal quadractic modulations with parameters $(r_d,N_d,N_d^{\text{WG}}) = (0.03,14,5)$ to form the optomechanical cavity.  The resulting cavity geometry is depicted in Figure \ref{fig:dim_diagram}(b).

\begin{figure}[t]
\begin{center}
\includegraphics[width=\columnwidth]{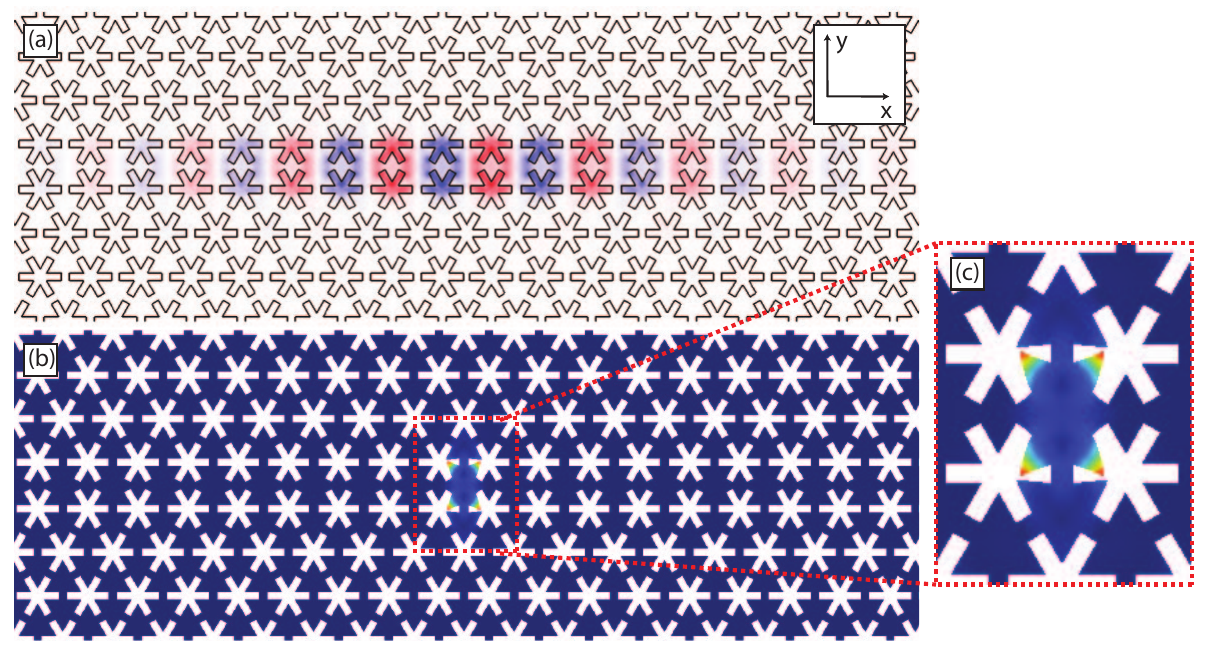}
\caption{Plots of the localized ultrahigh-$Q$ resonances of an optomechanical cavity formed in a Si snowflake thin-film substrate with parameters $(d,r,w,a)=(220,210,75,500)~\text{nm}$.  (a) Optical field ($E_y(\m r)$) and (b) magnitude of the mechanical displacement field ($\m Q(\m r)$). (c) Zoom-in of the mechanical displacement field.  In the optical field plot we show a snapshot in time of the $y$-polarization of the electric field, with red and blue indicating positive and negative values of the field, respectively.  In the mechanical mode plots, color indicates the magnitude of the displacement field (blue no displacement, and red large displacement), and the displacement of the structure has been exaggerated for viewing purposes.  The mechanical resonance is at a frequency of $\nu_m = 9.50~\text{GHz}$, and the optical mode at a wavelength of $\lambda_0 = 1.459~\mu\text{m}$. The defect cavity design, with parameters $(r_d,N_d,N_d^{\text{WG}}) = (0.03,14,5)$, is described in Fig.~\ref{fig:dim_diagram}. The lowest-order optomechanical coupling between the photon and phonon cavity resonances is calculated to be $g =2\pi \times 292~\text{kHz}$.}
\label{fig:final_cavity}
\end{center}
\end{figure}

As hoped, the resulting optomechanical cavity supports a localized fundamental mechanical mode of frequency $\nu_m = 9.5~\text{GHz}$ and a fundamental optical mode of frequency $\nu_o = 205.6~\text{THz}$, the latter corresponding to a free-space wavelength of $\lambda_0 = 1.459~\mu\text{m}$.  Plots of the FEM-simulated electric and mechanical displacement fields of both localized cavity resonances are shown in Fig.~\ref{fig:final_cavity}.  Owing to the extremely flat dispersion of the mechanical waveguide band, the cavity phonon resonance is localized almost entirely to the central unit cell of the cavity.  The resulting effective motional mass of this localized phonon is only $m_\text{eff} = 3.85~\text{fg}$ (where the maximum displacement of the mechanical vibration is used as the normal coordinate defining the mode\cite{Eichenfield2009d}).  The optomechanical coupling of photons and phonons in this cavity is calculated to be $g/2\pi = 292~\text{kHz}$. This value represents an enormous radiation pressure coupling of the optical and mechanical fields, being only a factor of two or so weaker than that found in the strongly-coupled zipper cavity~\cite{Eichenfield2009a} and only slightly below the approximate upper-bound calculated using equation (\ref{eqn:g_optimal}) of $g_\text{optimal}/2\pi = 382~\text{kHz}$.

The radiation-limited optical $Q$-factor was also computed for this cavity using perfectly-matched-layer radiation boundary conditions, and found to be limited to $Q \approx 5.1\times10^7$ due to out-of-plane radiation.  The corresponding mechanical $Q$-factor of the localized phonon resonances is $Q_m \gg 10^7$ for $40$ unit cells surrounding the cavity region. Unlike in the optical case, this mechanical $Q$ can be made arbitrarily large by increasing the number of unit cells surrounding the cavity due to the lack of an out-of-plane loss mechanism and a pseudo-bandgap in-plane. The important practical limiting factor in both the mechanical and optical $Q$ will of course be the presence of perturbations in a real fabricated structure. From Fig.~\ref{fig:waveguide_bands} we see that in the absence of perturbations breaking the $z$-mirror symmetry,  the mechanical radiation loss can be made effectively zero due to the complete lack of states to which the phononic resonance can couple. Perturbations breaking the $z$-mirror symmetry will however induce loss by coupling to odd vertical symmetry mechanical waveguide modes, colored green in Fig.~\ref{fig:waveguide_bands}.  Terminating the waveguide, i.e. transitioning back into the bulk snowflake crystal after some number of unit cells, eliminates this component of mechanical radiative loss as well. 

\section{Conclusions}

By introducing a new crystal structure, the ``snowflake'' structure, with photonic and phononic properties amenable to the formation of low-loss optical and acoustic resonances, we have shown that the on-chip control and interaction of photons and phonons may be possible in a realistic setting. This crystal structure, along with the waveguides and cavities proposed in this paper, allow for the unification of the quasi-2D photonic and phononic crystal systems. In the future, systems utilizing this unification may not only have wide applicability in classical optics~\cite{Safavi-Naeini2009}, but could also provide new ways to do quantum information science by allowing for a new class of hybrid quantum systems \cite{rabl-2009,wallquist-2009-137,Safavi-Naeini2009}.

\section*{Acknowledgements}  This work was funded through the NSF under EMT grant no. 0622246, MRSEC grant no. DMR-0520565, and CIAN grant no. EEC-0812072 through University of Arizona.

\end{document}